\documentclass[amsmath,amssymb,aps,floatfix,prl,reprint]{revtex4-1} 

\usepackage{graphicx}
\pdfoutput=1

\usepackage{bm}
\newcommand{\mb}{\mathbf}

\begin{document}
\title{Friction of water on graphene and hexagonal boron Nitride from \textit{ab initio} methods: very different slippage despite
very similar interface structures}
\author{Gabriele Tocci}
\affiliation{Thomas Young Centre}
\affiliation{London Centre for Nanotechnology, London WC1E 6BT, United Kingdom}
\affiliation{Department of Chemistry, University College London, London WC1H 0AJ, United Kingdom}
\author{Laurent Joly}
\affiliation{Institut Lumi\`ere Mati\`ere, UMR5306 Universit\'e Lyon 1-CNRS, Universit\'e de Lyon 69622 Villeurbanne, France}
\author{Angelos Michaelides}
\affiliation{Thomas Young Centre}
\affiliation{London Centre for Nanotechnology, London WC1E 6BT, United Kingdom}
\email{angelos.michaelides@ucl.ac.uk}

\begin{center}
\begin{abstract}
Friction is one of the main sources of dissipation at liquid water/solid interfaces. Despite recent progress,
a detailed understanding of water/solid friction in connection with the structure and energetics of the solid surface is lacking.
Here we show for the first time that \textit{ab initio} molecular dynamics can be used to unravel the connection between
the structure of nanoscale water and friction for liquid
water in contact with graphene and with hexagonal boron nitride. We find that whilst the interface
presents a very similar structure between the two sheets,
the friction coefficient on boron nitride is $\approx 3$ times larger than that on graphene.
This comes about because of the greater corrugation of the energy landscape on boron nitride arising from
specific electronic structure effects.
We discuss how a subtle dependence of the friction on the atomistic details of a surface, that is
not related to its wetting properties, may have a significant impact on the
transport of water at the nanoscale, with implications for the development of membranes for desalination
and for osmotic power harvesting.
\end{abstract}
\end{center}
\thispagestyle{empty}
\maketitle

Nanofluidics is an exciting field that offers alternative and
sustainable solutions to problems relating to energy conversion, water filtration and 
desalination~\cite{BNNT_bouquet,Holt19052006,Majumder2005,Hummer2001,graphene_membrane_geim,Logan_membrane_power_generation,Bocquet2010,Desalination_acsnano,striolo_langmuir}.
Miniaturization towards nanofluidic devices inevitably leads to an enhanced influence of
surface and interface properties as opposed to those of the bulk.
Friction is the most important interface property that
limits fluid transport at the nanoscale, and its understanding is therefore crucial
for the design of more efficient membranes, nanotubes and pores that exhibit low liquid/solid friction.
The behavior of  liquid flow at
scales on the order of a few tens of nanometres departs from continuum
fluid dynamics and desirable transport properties emerge at such small scales~\cite{Bocquet2007}.
For instance, carbon nanotubes have
a very high water permeability as compared to the prediction of macroscopic fluid dynamics~\cite{Holt19052006}. Further,
a vanishing friction has been found, giving rise to superlubric behavior
of water chains inside tubes of sub-nanometre radii~\cite{laurent_friction}.

Besides carbon, boron nitride (BN) nanostructures have recently been explored for the development
of nanofluidic devices for fast water transport and efficient power
generation~\cite{BNNT_bouquet,water_BN_MD_won,MD_aluru_jpcc}.
Recent interest has been fueled by the demonstration that salinity concentration gradients
across BN nanotube membranes can leed to the generation of very large
electric currents~\cite{BNNT_bouquet}.
It has also very recently been shown that there is a very large inter-layer friction between
in multiwalled BN nanotubes~\cite{BNNT_bouquet_friction}, as opposed to the
superlubric behavior of the (homopolar) carbon nanotubes~\cite{super_lubricity_CNT}.
This suggests that the frictional properties of BN and C nanostructures might
be quite different.
However, to the best of our knowledge there has been no attempt to measure
or compute the friction of water at the interface with BN sheets or nanotubes.
Given that \textit{ab initio} results have shown very similar contact angles of
water droplets on graphene and BN sheets~\cite{water_contact_angle_gra_bn},
it remains to be seen if transport properties on these two systems are also similar.

The rise of the atomic force microscope and the surface force apparatus has advanced our understanding of nanoscale
liquid/solid friction substantially~\cite{tosatti_colloquium}. Yet,
it remains extremely hard to relate friction and dynamics to structure and wetting of solid surfaces from experiment.
This is especially true for the case of graphene where there is even controversy
over the water contact angle (see \textit{e.g.} Refs.~\cite{wetting_break_down,wetting_transparency}).
Molecular simulations, and especially force field molecular dynamics,
have proved extremely useful in investigating structural and dynamical properties of confined liquids
elucidating molecular level information that is challenging for experiments to obtain.
Because of the dependence of simulation results on the parametrization of force fields,
it is interesting to investigate the dynamics of interfacial water using electronic structure methods
so that two different materials such as graphene and BN can be compared on an equal footing.
Accordingly, \textit{ab initio} molecular dynamics (AIMD) offers an interesting alternative that has
been widely used to study complex liquid/solid interfaces (see \textit{e.g.} Refs.~\cite{Galli_water_graphene,tocci2014ZnO,water_contact_angle_gra_bn}).
Although transport properties in liquids have been computed before using AIMD
(see \textit{e.g.} Refs.~\cite{kuhne_viscosity,alfe_viscosity,feibelman_viscosity}), we show for the first time
that it is possible to compute a converged friction coefficient from AIMD.

In this work we compare the structure and dynamics of water in contact with BN and carbon nanostructures from AIMD.
Specifically, we study liquid water in contact with graphene
and hexagonal BN sheets, which is relevant to understanding
flow at membranes based on these materials and also inside large nanotubes~\cite{note_graphene}.
We find a striking similarity between the structure of water in contact with the two sheets.
Nevertheless, there is a three-fold increase in the friction coefficient on BN
because of a more corrugated free energy surface on BN compared to graphene.
This work illustrates the complexity of nanoscale friction where subtle electronic effects,
with no detectable consequences on the structure of the liquid, may however have a dramatic
impact on water transport at the nanoscale.

\begin{figure}[!h]
{\includegraphics[width=0.5\textwidth]{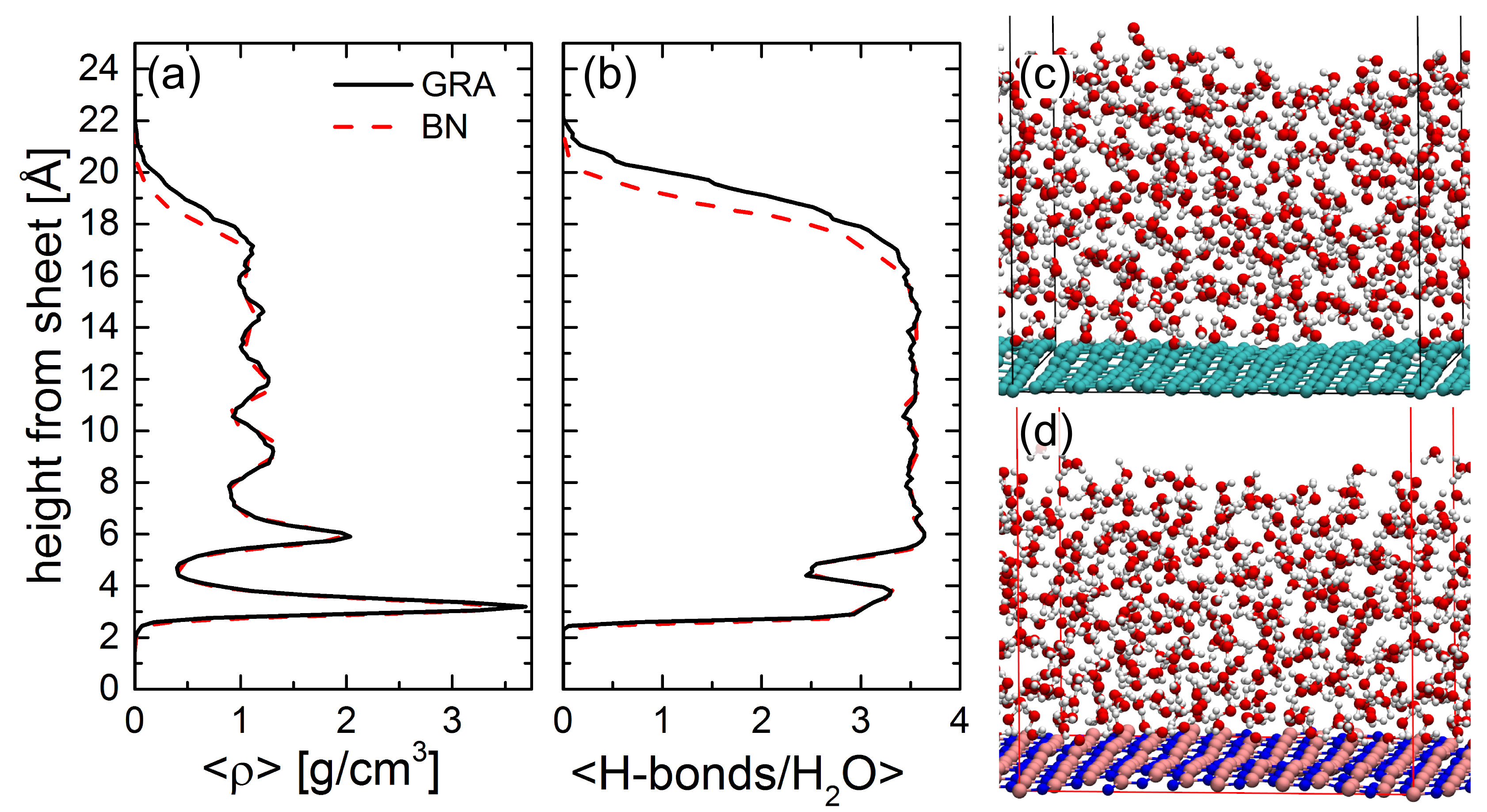}
\caption{Structure of the liquid water film on graphene (GRA) and a single sheet of hexagonal BN.
Average density profile $\langle \rho \rangle$ (a) and number of hydrogen bonds $\langle$H-bonds$/$H$_2$O$\rangle$  (b)
as a function of the height from the sheet. In (b) the geometric criterion from Luzar and Chandler was used to define a H-bond~\cite{luzar_chand}.
Snapshots of the liquid film on GRA (c) and on BN (d). In (c) and (d) O and H atoms are coloured in red and white,
while C atoms are in light blue, and B and N atoms are in pink and purple, respectively.
The liquid film structure is essentially the same on both sheets, only on BN the film is slightly thinner because
the BN unit cell is $\approx 3.5\%$ larger than  that of graphene.}
\label{fig:AIMD_dens}}
\end{figure}

We performed a series of extensive AIMD simulations of a
 thin liquid water film on graphene and on a single layer of hexagonal BN.
Full details of the computational set-up can be found in the Supporting Information (SI).
However, the key features are that we used the CP2K~\cite{cp2kvond} code to perform
AIMD simulations in the Born-Oppenheimer approximation with the electronic structure computed at the
density functional theory (DFT) level using the optB88-vdW exchange-correlation functional~\cite{jiri_molecules,jiri_solids}.
The optB88-vdW density functional has been shown to predict
accurate interlayer binding and interlayer distances in graphite and bulk hexagonal BN~\cite{graziano_vdw_gra_bn},
to give a good description of the structure of bulk liquid water~\cite{Galli_water_jctc}
and also to correctly describe the relative stability of
water-ice structures on metals compared to the bulk ice lattice energy~\cite{javi_vdw_prl}.
The water/graphene and water/BN sheets have been modelled using $6\times 10$ orthorhombic cells about $25\times 25$ {\AA}$^2$ wide,
with $\approx 20$ {\AA} thick liquid water films.
There is a vacuum gap between the liquid-vacuum interface and the next periodic image of $\approx 15$ {\AA}.
Each film contains 400 water molecules and in total each system consists of 1440 atoms. Upon these systems we
performed 40 ps long AIMD simulations with the last 35 ps of each trajectory used for analysis.
As an illustration snapshots from the AIMD simulations for the water/graphene and water/BN interfaces
are shown in Fig.~\ref{fig:AIMD_dens}(c) and (d), respectively.
The AIMD simulations are in the canonical ensemble close to room temperature ($330$ K).
We also performed a number of additional AIMD and force field MD simulations to
explore the sensitivity of our results to the use of a different ensemble
and to issues such as finite size effects, different initial conditions and time scales.
Overall, the results of these tests, which can be found in the SI,
agree with the results presented herein.

We begin our analysis by illustrating in Fig.~\ref{fig:AIMD_dens}(a) the planar average density profile $\langle \rho\rangle$
as a function of the height from the sheets. The graphene and BN sheets
exhibit oscillations in the heights of the atoms within them of up to
1.2 {\AA}. To account for this, we compute the height of an atom in the liquid water overlayer as
the height difference to the closest atom in the sheet.
By computing the density profile in this way we fully account for the layering of the liquid film,
which would otherwise be partially hidden behind the oscillations of the sheets.
The two density profiles on graphene and BN overlap for most of the film height, which is the first signature of
the apparent similarity of the structure of liquid water on the two sheets. Perturbation from the bulk liquid induced by the surface 
is most significant within the first $10$ {\AA} of the surface. This is
consistent with previous reports on liquid water/solid interfaces
(see \textit{e.g.} Refs.~\cite{Galli_water_graphene,tocci2014ZnO,liq_wat_surfaces_rossky,striolo_water_graphene_polarizability}),
which extends in this case to about 8 {\AA} from the sheets. Within this region there are two evident peaks for
each of the two density profiles. The first at a height of about 3.0 {\AA} hits a density maximum of $\approx 3.7 $ g/cm$^3$.
After this first peak there is a depletion of water with a minimum at about 4.5 {\AA} with a density of 0.4 g/cm$^3$. We define the contact layer
as the region delimited by this minimum in the density profile. At $\approx 6.0$
{\AA} the second peak appears with a density of $\approx 2$ g/cm$^3$.
Further away from the sheets, density oscillations are gradually suppressed
and the density of liquid water is recovered with an average value of $1.12\pm 0.16 $ g/cm$^3$.
Finally, the density decays as it is characteristic of the liquid-vapor interface~\cite{wat_kuo_sci}.
The decay in the density on BN starts off at about 17 {\AA}, slightly before than on graphene
simply because the BN unit cell is $\approx 3.5\%$ larger.

As further confirmation of the striking similarity between the structure of liquid water on graphene and BN,
we compare the profile of the average number of H-bonds per molecule, $\langle$H-bonds$/$H$_2$O$\rangle$ in Fig.~\ref{fig:AIMD_dens}(b).
Also in this case the two curves essentially lie on top of each other for
all of the film height apart from the liquid-vapour interface region.
The steep increase in the  $\langle$H-bonds$/$H$_2$O$\rangle$ starting at about 2 {\AA} shows that already in the contact layer
water molecules engage in a large number of H-bonds of $ \approx 3 $/H$_2$O. Shortly after this initial increase, there is a drop
to $\approx 2.5$ H-bonds$/$H$_2$O corresponding to the depletion region around 4.5 {\AA} from the sheet. 
The pronounced fluctuations in the number of water
molecules  in this region partly penalizes H-bonding with neighboring waters. After the depletion region, at a height of 5 {\AA},
the number the $\langle$H-bonds$/$H$_2$O$\rangle$ rises to the bulk value of $\approx 3.5$. It remains constant until the liquid-vapour region
is approached, at a height of 17 {\AA}, where there is a rapid decay
 to zero in the $\langle$H-bonds$/$H$_2$O$\rangle$ within 3 to 4 {\AA}.
Finally, other structural characteristics of the two systems exhibit
striking similarities, such as the orientations of the water molecules within the films (see Fig.~S2 in the SI).

\begin{figure}[!h]
{\includegraphics[width=0.5\textwidth]{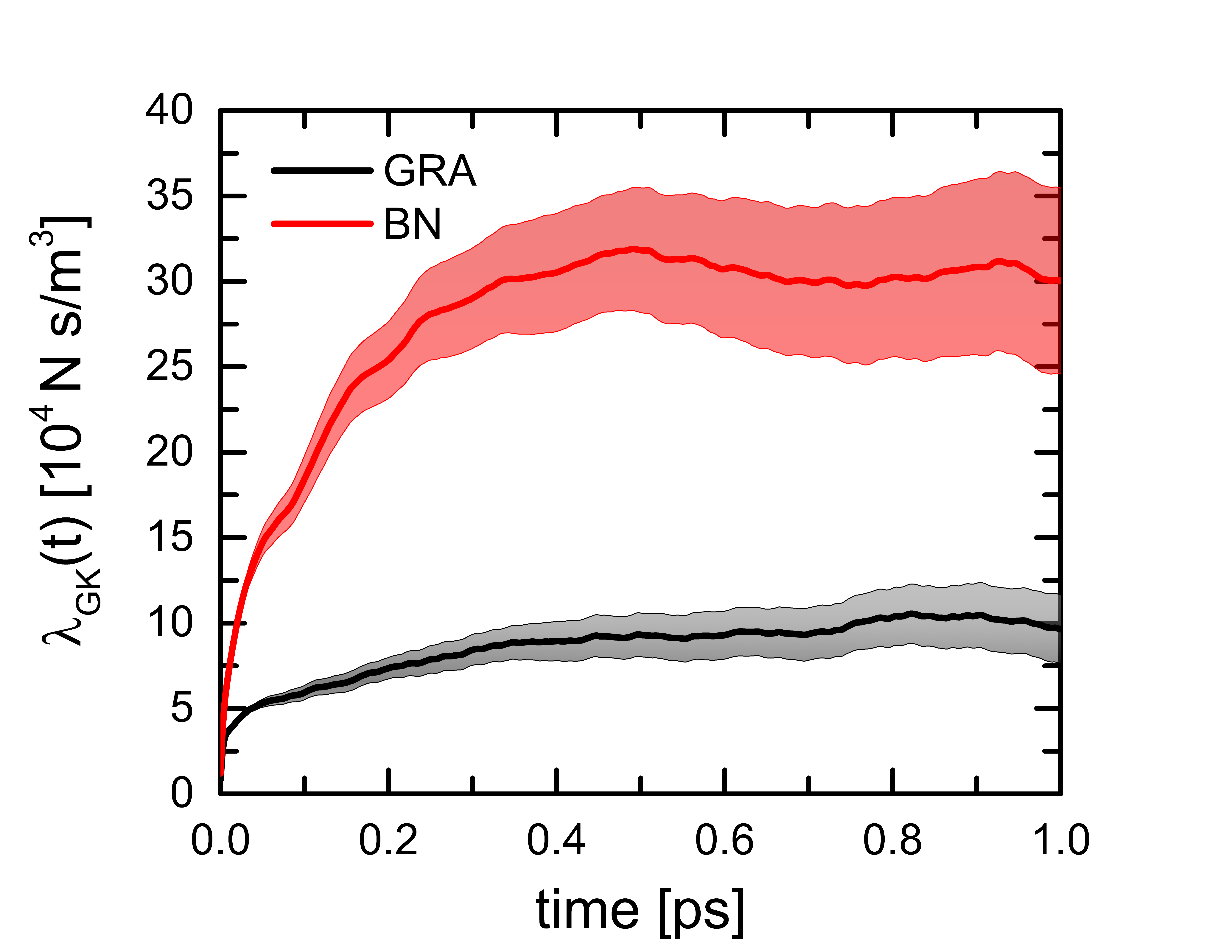}
\caption{Comparison between the Green-Kubo estimate of the friction coefficient of liquid water on graphene (GRA) and on BN.
The shaded areas represent the uncertainties obtained by performing a block average.
The friction coefficient $\lambda$ is given by the plateau value at long times.
There is an evident increase in the friction coefficient on BN.}
\label{fig:AIMD_fric}}
\end{figure}

Having compared the structure of liquid water on the two sheets we now turn our discussion to investigate its dynamics.
Specifically, we focus on the friction coefficient $\lambda$,
defined as the ratio between the friction force parallel to the sheet $\mb{F_p}$ per unit area ($\mathcal{A}$)
and $v_{slip}$, the velocity jump at the interface~\cite{Bocquet2007}, namely $\lambda=\mb{F_p}/(\mathcal{A}$\,$v_{slip})$.
In the framework of linear response theory, $\lambda$ can be obtained from the
equilibrium fluctuations of the friction force, using a Green-Kubo relation~\cite{Bocquet1994,Bocquet2013}: 
\begin{equation}\label{eq:lambda}
\lambda= \lim\limits_{\substack{t \to \infty}} \lambda_{GK} (t) ,
\end{equation}
with
\begin{equation}\label{eq:lambda_t}
\lambda_{GK} (t) = \frac{1}{2 \mathcal{A} k_\text{B}T} \int_0^t \langle \mb{F_p}(t') \cdot \mb{F_p}(0) \rangle\,\mathrm{d}t'
\end{equation}
where $k_\text{B}$ is the Boltzmann constant, $T$ the temperature and the factor of $1/2$ comes from the
averaging over the two spatial dimensions parallel to the sheets.
We show in Fig.~\ref{fig:AIMD_fric} $\lambda_{GK}(t)$ for the case of water on graphene and on BN,
which for sufficiently long time intervals (here approximately $>0.3$ ps) plateaues to the value of $\lambda$.
The key result is that the friction coefficients on the two sheets differ significantly, with the friction on
BN being approximately 3 times larger than on graphene.
Specifically, while $\lambda=(9.6 \pm 2.0) \times 10^4$\,N\,s\,m$^{-3}$ for graphene, we obtain a value on
BN of  $\lambda=(30.0 \pm 5.4)  \times 10^4$\,N\,s\,m$^{-3}$.
 A number of tests using both AIMD and force field MD,
ensured that our results on the friction coefficients are converged and
that the increase of about three times in the
friction on BN compared to graphene is consistently observed,
regardless of the type of ensemble used, time scales, or different system sizes.
The results of these tests are presented in Fig.~S3 in the SI.

Whilst liquid/solid friction is the relevant microscopic property that quantifies the
dynamics of a fluid at the nanoscale, a length scale characteristic of the flow is often measured experimentally.
This is the so called slip length $b$, defined as the distance
relative to the surface where the linear extrapolation of the tangential flow velocity vanishes.
We can relate the slip length $b$ to the friction coefficient via the shear viscosity of
bulk liquid water $\eta$: $b=\eta/\lambda$~\cite{Bocquet2007}.
Using the experimental bulk water viscosity $\eta = 10^{-3}$\,Pa\,s, the corresponding slip lengths
for graphene and BN are  $ 10.4\pm 2.2$ and $3.3 \pm 0.6$\,nm, respectively.
Overall water slippage on graphene and BN is characteristic of hydrophobic surfaces
with a low friction coefficient, while on hydrophilic surfaces such as mica, silicon or graphene oxide
slippage is significantly inhibited with sub-nm slip lengths~\cite{Huang2008b,ncomm_young_friction}.

\begin{figure}[!h]
{\includegraphics[width=0.5\textwidth]{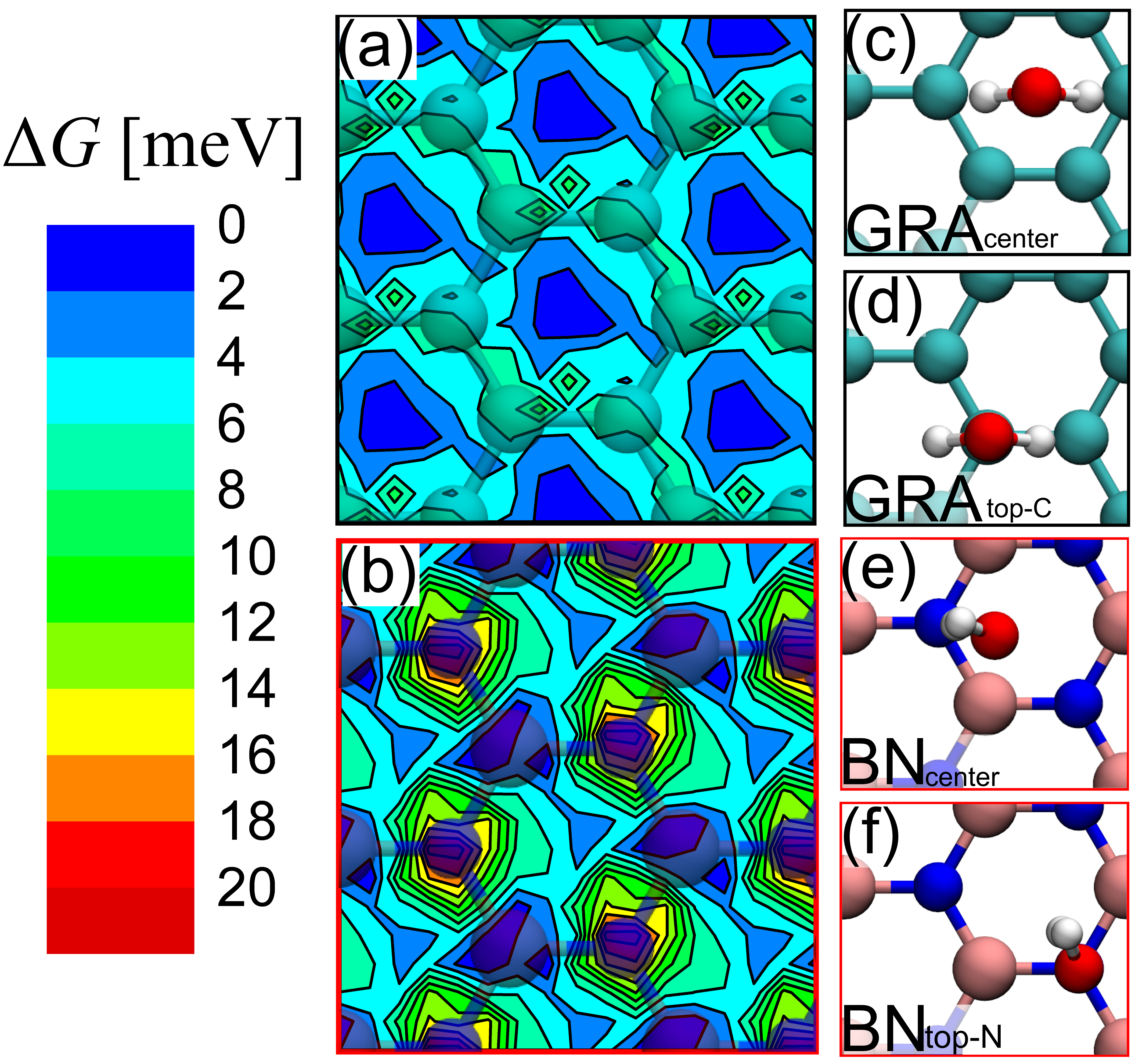}
\caption{Free energy profile of water within the contact layer of the liquid projected onto
the graphene (a) and BN (b) primitive unit cells.
Although the free energy profiles are relatively smooth,
a larger corrugation is present on BN and some differences are also observed
in the topography. Transparent C and BN atoms are superimposed on the contour plots.
Most stable (c) and less stable (d) configuration for a single water monomer adsorbed on graphene.
Most stable (e) and less stable (f) configuration for a single water monomer adsorbed on BN.
Only a small part of the unit cells used for the calculation of the monomer adsorption calculations
is shown in figures  (c) to (f).}
\label{fig:Delta_F}}
\end{figure}

To rationalize the difference in the two friction coefficients in terms of structural and energetic contributions,
we compute the free energy profile of water within the contact layer
$\Delta G(x,y)$, defined as $\Delta G (x,y) = - k_B T \ln P_O (x,y)$. Here, $P_O (x,y)$ is the spatial
probability distribution function of the O atom of water within the
contact layer at a point $(x,y)$ projected onto the primitive graphene and BN unit cells.
A similar approach based on the 2-D spatial probability distribution function has been used
previously to understand water slippage on model MgO surfaces~\cite{striolo_PNAS}.
Fig.~\ref{fig:Delta_F} illustrates the free energy profiles resulting from this analysis.
We notice the very small energy scale within $k_B T$ at room temperature indicating a mobile contact layer.
Although,  the average liquid structures in the two systems
are very similar, the free energy profiles for the motion of water in the contact layer
exhibit some clear differences.
First, the free energy minimum on graphene is around the hollow site
with the maximum around the top-C site, in agreement with previous work~\cite{Galli_water_graphene}.
The minimum on BN  is for the oxygens to sit around the B-site, 
as well as on the hollow site, while the maximum is on the N site.
Second, and more importantly, the BN free energy profile (Fig.~\ref{fig:Delta_F}(b))
is more corrugated than that of graphene: the maximum corrugation of the free energy 
is only 13 meV on graphene, but it is 21 meV on BN.
Although we are discussing very small energies, 
this $\approx 60\%$ increase in the corrugation
is observed consistently in all our various AIMD simulations of these systems.
Indeed, the corrugation is already converged if the free energy profile is extracted
from half the AIMD trajectory, as opposed to the full trajectory (see Fig.~S5).
As we now discuss, this increased corrugation
is the main reason for the observed increase in the friction coefficient on BN.

It is interesting that such small energy differences, less than 10 meV can contribute to
a three fold increase in the friction coefficient and it is worth looking into this more in detail.
It is known that the leading term in the friction depends quadratically on the corrugation of the
potential energy surface felt by the water in the contact layer $\Delta V$,
such that  $\lambda \sim \Delta V^2$ (see \textit{e.g.}~\cite{Bocquet2010,krim_friction_rev}).
We can approximate the corrugation of the potential
energy  $\Delta V$ with the corrugation in $\Delta G$ as obtained from
the maximum value in free energy profiles in Fig.~\ref{fig:Delta_F}, such that $\lambda\sim \Delta G ^2$.
To test the validity of this scaling relation we computed the ratio between the
friction of water on BN and on graphene, $ \lambda_{BN}/ \lambda_{GRA}$ and
compared it with the ratio between the square of the free energy extracted from Fig.~\ref{fig:Delta_F}(a) and (b).
From the ratio between the friction coefficients we obtain  $ \lambda_{BN}/ \lambda_{GRA} = 3.1 \pm 0.8$.
The corrugation of the free energy on BN is 21 meV, while it is 13 meV on graphene, 
such that the ratio $(\Delta G_{BN}/\Delta G_{GRA})^2\approx 2.6$,
is  within the error of the ratio between the two values of the friction.

The free energy corrugation depends on the atomic and
 electronic structure of the surface and on the H-bonds that form at a specific interface.
It will differ but be related
 to the monomer potential energy surface.
By performing an extensive
series of calculations (about 4000 wave function optimizations per sheet)
we mapped the potential 
energy surface of a water monomer on the sheets (see Fig.~S1),
to explore if the potential of 
a water monomer on BN is more corrugated than that on graphene.
This confirms that the free energy of the water contact layer on BN is more corrugated than that on graphene
because the potential of the water monomer is also more corrugated.

From the analysis of the different structures of a water monomer on the two sheets,
we have identified the main sites that give rise to
the most and least stable configurations of water on graphene and BN.
The most stable structure on graphene is for water on the center of the hexagon
 with its dipole pointing down towards the sheet
(Fig.~\ref{fig:Delta_F}(c)). On BN water preferentially adsorbs in the center of the hexagon with
the dipole moment parallel to the sheet and with one of the O--H bonds pointing towards the N atom (Fig.~\ref{fig:Delta_F}(e)).
Upon considering water monomer adsorption at less favorable sites (see Figs.~\ref{fig:Delta_F}(d) and (f)) we find that
indeed the potential energy surface of the water monomer on BN is more corrugated than that on graphene (see Table~S1).
To understand the reason for the different corrugation between the two sheets, we performed a decomposition of
the interaction energies of several water structures on the sheets at the more stable sites
(on the center of the hexagon of graphene and BN) and at less stable sites (on the top-C and top-N site), as shown in Table~S1.
Although there is a certain level of arbitrariness to most decomposition schemes, we
have used an established approach to gain qualitative and semi-quantitative
insight into our adsorption systems (see \textit{e.g.} Refs.~\cite{javi_vdw_prl,graziano_vdw_gra_bn}).
This analysis reveals that
van der Waals dispersion forces and local-correlation effects do not contribute to
the corrugation of the graphene and BN sheets. On the other hand,
the interactions arising from exchange, electrostatic and electron kinetic energy terms
are overall more pronounced on BN and provide the reason for the increased corrugation,
and thus for the 3-fold increase in the friction.
Finally, we note that even though very small energy differences are
being reported for the corrugation of the free energy profile,
our AIMD simulations are long enough that the free energy profiles are converged even if they
are extracted over half the trajectory (see Fig.~S5). Also,
these differences are robust with respect to the accuracy of the
underlying potential energy surface (see Fig.~S1) and they are consistent with the
monomer interaction energies computed in Table~S1 using different codes and
exchange-correlation functionals. 

Since this is the first AIMD study of water slippage at either of the two sheets examined,
we wish to compare our results with the existing literature.
In the absence of experiments on graphene or BN we compare our results to those
obtained from atomic force microscopy measurements of
water droplets on atomically smooth highly ordered pyrolitic graphite, which have given slip lengths
of between 8 and 12 nm~\cite{mali_friction_water_graphite,ncomm_young_friction},
in good agreement with our value for the slip length of $10.4 \pm 2.2$ nm.
Although on graphene water transport may differ from that on graphite due to \textit{e.g.}
screening effects that may influence the binding on the two systems~\cite{tkatc_natur_comm_2013},
we do not expect the friction to change between graphene and graphite by more than a few percent.

Compared to previous simulation studies, we notice that
there is some uncertainty over the slippage of water on graphene obtained from force field MD,
with values for the slip length between 1 and 80\,nm~\cite{Kannam_JCP},
and our AIMD result falls in this somewhat large range.
The value of the friction on graphene
obtained from our force field MD is $\approx 3 \times 10^4$\,N\,s\,m$^{-3}$ (see Fig.~S3(b)),
about 3 times smaller than our AIMD value of $(9.6 \pm 2.0) \times 10^4$\,N\,s\,m$^{-3}$.
This is because the optB88-vdW functional predicts a larger corrugation of the free energy
compared to the force field used~\cite{Werder2003}(see Fig.~S4).
Since optB88-vdW overestimates the absolute adsorption energy of a water monomer on graphene
when compared to  benchmark diffusion Monte Carlo and Random Phase Approximation
results~\cite{ma_monomer_on_gra}, we cannot expect optB88-vdW to capture the corrugation of the
free energy with absolute precision. However, 
as explained earlier, because this functional successfully
captures various other properties of water, graphite, BN and water at interfaces we expect it
to predict the correct relative water/graphene and water/BN interaction strengths
and hence the correct increase in the friction on BN.
Since the increase in friction is not captured
by force field molecular dynamics (see Fig.~S3(b)),
the findings here stress the importance of accounting
for electronic structure effects when investigating transport properties
at complex liquid/solid interfaces. Nevertheless,
it is possible that improving the description of the force fields to include polarizable models
and partial charges on B and N may reproduce the observed friction increase.
For instance, it has been found that including polarization effects in force field molecular dynamics studies
has an effect on the diffusion of liquid water on charged graphene~\cite{striolo_water_graphene_polarizability}.

Finally, there has been increasing interest in connecting wetting properties to
the friction coefficient (see \textit{e.g.} Ref.~\cite{Huang2008b}) and a relation
between the slip length and the contact angle has been found to hold for a wide number of liquid/solid interfaces.
Here we have seen that the structure of liquid water on graphene and on BN  is strikingly similar and previous \textit{ab initio}
work reported also contact angles of $86^{\circ}$ and $87^{\circ}$ on graphene and BN, respectively~\cite{water_contact_angle_gra_bn}.
Yet, friction is about three times larger on BN, highlighting that
a simple dependence on the wetting properties cannot be established for these two systems.
Instead, we have demonstrated a dependence of the friction on the free energy of the water contact layer.
In systems like graphene and BN, where the corrugation of the potential felt
by the water is not directly proportional to the water/solid interaction strength,
the free energy profile provides a closer estimate of the potential energy landscape corrugation.
Possible other examples are carbon nanotubes, which like graphene, have been shown to depart from the scaling law
that relates the friction to the water contact angle~\cite{laurent_friction}; and most likely BN nanotubes and other
van der Waals layered materials~\cite{vdw_hetero_structures}.
With the aim of designing nanofluidic devices which exhibit frictionless fluid transport,
materials may be engineered to favor a smooth potential energy landscape independently of the fluid/solid interaction strength.
In this manner for instance, the friction may be tailored whilst maintaining the same wetting properties.
We have shown that because of electronic structure effects, friction is larger on
BN. Limiting the magnitude of such interactions between the liquid and the substrate
using for instance homopolar surfaces and avoiding the presence of H-bonds
is beneficial for the design of smooth interfaces.
Further,  compressive or tensile stress can be applied to a substrate to favor a smooth contact layer~\cite{ming_ma_PRE_2011}.

In conclusion, we have reported on extensive \textit{ab initio}
molecular dynamics studies of liquid water on graphene and BN.
In so doing we have tried to bridge the gap between the molecular structure and energetics
of water on layered materials and the complex transport of water at
the nanoscale fully from \textit{ab initio} methods.
A comparison between the two systems reveals that while the structure of the liquid film is very
similar, slight differences in the water contact layer
not related to the wetting properties of the interfaces
give rise to a remarkably different water slippage on the two sheets.
The three-fold friction increase on BN is induced by a larger
corrugation of the energy landscape compared to graphene, because of more pronounced electrostatic and exchange interactions.
Overall, this work paves the way for the study of transport properties
in yet more complex liquid water/solid interfaces using
\textit{ab initio} methods. For example, systems where water is liable to
dissociate or where ions, defects, or external electric fields
are present could all be examined.
We also hope that our work will stimulate the
study of nanoscale water friction on BN-nanostructures and other layered
materials using \textit{e.g.} atomic force microscopy~\cite{ncomm_young_friction}.
\section{Supplementary Information}
Further computational details and tests on the structure, 
energetics and friction coefficient of the water/graphene and water/BN interfaces.
\section{Acknowledgements}
We are grateful to Lyderic Bocquet for encouraging us to carry out this work.
The early stages of this work were supported by the FP7 Marie Curie Actions of 
the European Commission, via the Initial Training Network SMALL (MCITN-238804).
LJ is supported by the French Ministry of Defense through the project
DGA ERE number 2013.60.0013.00.470.75.01, and by the ``labex iMUST''
through the ``mobilit\'e out'' program.
A. M. is supported by the European Research Council (HeteroIce project) and the Royal Society
through a Wolfson Research Merit Award.
We are also grateful for computational resources to the London Centre for
Nanotechnology and to the UK's HPC Materials Chemistry
Consortium, which is funded by EPSRC (EP/F067496), and to the UKCP consortium, EP/F036884/1
for access to HECToR and Archer.
\bibliography{water_gra_bn_bib}

\begin{thebibliography}{49}%
\makeatletter
\providecommand \@ifxundefined [1]{%
 \@ifx{#1\undefined}
}%
\providecommand \@ifnum [1]{%
 \ifnum #1\expandafter \@firstoftwo
 \else \expandafter \@secondoftwo
 \fi
}%
\providecommand \@ifx [1]{%
 \ifx #1\expandafter \@firstoftwo
 \else \expandafter \@secondoftwo
 \fi
}%
\providecommand \natexlab [1]{#1}%
\providecommand \enquote  [1]{``#1''}%
\providecommand \bibnamefont  [1]{#1}%
\providecommand \bibfnamefont [1]{#1}%
\providecommand \citenamefont [1]{#1}%
\providecommand \href@noop [0]{\@secondoftwo}%
\providecommand \href [0]{\begingroup \@sanitize@url \@href}%
\providecommand \@href[1]{\@@startlink{#1}\@@href}%
\providecommand \@@href[1]{\endgroup#1\@@endlink}%
\providecommand \@sanitize@url [0]{\catcode `\\12\catcode `\$12\catcode
  `\&12\catcode `\#12\catcode `\^12\catcode `\_12\catcode `\%12\relax}%
\providecommand \@@startlink[1]{}%
\providecommand \@@endlink[0]{}%
\providecommand \url  [0]{\begingroup\@sanitize@url \@url }%
\providecommand \@url [1]{\endgroup\@href {#1}{\urlprefix }}%
\providecommand \urlprefix  [0]{URL }%
\providecommand \Eprint [0]{\href }%
\providecommand \doibase [0]{http://dx.doi.org/}%
\providecommand \selectlanguage [0]{\@gobble}%
\providecommand \bibinfo  [0]{\@secondoftwo}%
\providecommand \bibfield  [0]{\@secondoftwo}%
\providecommand \translation [1]{[#1]}%
\providecommand \BibitemOpen [0]{}%
\providecommand \bibitemStop [0]{}%
\providecommand \bibitemNoStop [0]{.\EOS\space}%
\providecommand \EOS [0]{\spacefactor3000\relax}%
\providecommand \BibitemShut  [1]{\csname bibitem#1\endcsname}%
\let\auto@bib@innerbib\@empty
\bibitem [{\citenamefont {Siria}\ \emph {et~al.}(2013)\citenamefont {Siria},
  \citenamefont {Poncharal}, \citenamefont {Biance}, \citenamefont {Fulcrand},
  \citenamefont {Blase}, \citenamefont {Purcell},\ and\ \citenamefont
  {Bocquet}}]{BNNT_bouquet}%
  \BibitemOpen
  \bibfield  {author} {\bibinfo {author} {\bibfnamefont {A.}~\bibnamefont
  {Siria}}, \bibinfo {author} {\bibfnamefont {P.}~\bibnamefont {Poncharal}},
  \bibinfo {author} {\bibfnamefont {A.-L.}\ \bibnamefont {Biance}}, \bibinfo
  {author} {\bibfnamefont {R.}~\bibnamefont {Fulcrand}}, \bibinfo {author}
  {\bibfnamefont {X.}~\bibnamefont {Blase}}, \bibinfo {author} {\bibfnamefont
  {S.~T.}\ \bibnamefont {Purcell}}, \ and\ \bibinfo {author} {\bibfnamefont
  {L.}~\bibnamefont {Bocquet}},\ }\href@noop {} {\bibfield  {journal} {\bibinfo
   {journal} {{Nature}}\ }\textbf {\bibinfo {volume} {494}},\ \bibinfo {pages}
  {455} (\bibinfo {year} {2013})}\BibitemShut {NoStop}%
\bibitem [{\citenamefont {Holt}\ \emph {et~al.}(2006)\citenamefont {Holt},
  \citenamefont {Park}, \citenamefont {Wang}, \citenamefont {Stadermann},
  \citenamefont {Artyukhin}, \citenamefont {Grigoropoulos}, \citenamefont
  {Noy},\ and\ \citenamefont {Bakajin}}]{Holt19052006}%
  \BibitemOpen
  \bibfield  {author} {\bibinfo {author} {\bibfnamefont {J.~K.}\ \bibnamefont
  {Holt}}, \bibinfo {author} {\bibfnamefont {H.~G.}\ \bibnamefont {Park}},
  \bibinfo {author} {\bibfnamefont {Y.}~\bibnamefont {Wang}}, \bibinfo {author}
  {\bibfnamefont {M.}~\bibnamefont {Stadermann}}, \bibinfo {author}
  {\bibfnamefont {A.~B.}\ \bibnamefont {Artyukhin}}, \bibinfo {author}
  {\bibfnamefont {C.~P.}\ \bibnamefont {Grigoropoulos}}, \bibinfo {author}
  {\bibfnamefont {A.}~\bibnamefont {Noy}}, \ and\ \bibinfo {author}
  {\bibfnamefont {O.}~\bibnamefont {Bakajin}},\ }\href {\doibase
  10.1126/science.1126298} {\bibfield  {journal} {\bibinfo  {journal}
  {Science}\ }\textbf {\bibinfo {volume} {312}},\ \bibinfo {pages} {1034}
  (\bibinfo {year} {2006})}\BibitemShut {NoStop}%
\bibitem [{\citenamefont {Majumder}\ \emph {et~al.}(2005)\citenamefont
  {Majumder}, \citenamefont {Chopra}, \citenamefont {Andrews},\ and\
  \citenamefont {Hinds}}]{Majumder2005}%
  \BibitemOpen
  \bibfield  {author} {\bibinfo {author} {\bibfnamefont {M.}~\bibnamefont
  {Majumder}}, \bibinfo {author} {\bibfnamefont {N.}~\bibnamefont {Chopra}},
  \bibinfo {author} {\bibfnamefont {R.}~\bibnamefont {Andrews}}, \ and\
  \bibinfo {author} {\bibfnamefont {B.~J.}\ \bibnamefont {Hinds}},\ }\href@noop
  {} {\bibfield  {journal} {\bibinfo  {journal} {Nature}\ }\textbf {\bibinfo
  {volume} {438}},\ \bibinfo {pages} {44} (\bibinfo {year} {2005})}\BibitemShut
  {NoStop}%
\bibitem [{\citenamefont {Hummer}\ \emph {et~al.}(2001)\citenamefont {Hummer},
  \citenamefont {Rasaiah},\ and\ \citenamefont {Noworyta}}]{Hummer2001}%
  \BibitemOpen
  \bibfield  {author} {\bibinfo {author} {\bibfnamefont {G.}~\bibnamefont
  {Hummer}}, \bibinfo {author} {\bibfnamefont {J.~C.}\ \bibnamefont {Rasaiah}},
  \ and\ \bibinfo {author} {\bibfnamefont {J.~P.}\ \bibnamefont {Noworyta}},\
  }\href@noop {} {\bibfield  {journal} {\bibinfo  {journal} {Nature}\ }\textbf
  {\bibinfo {volume} {414}},\ \bibinfo {pages} {188} (\bibinfo {year}
  {2001})}\BibitemShut {NoStop}%
\bibitem [{\citenamefont {Nair}\ \emph {et~al.}(2012)\citenamefont {Nair},
  \citenamefont {Wu}, \citenamefont {Jayaram}, \citenamefont {Grigorieva},\
  and\ \citenamefont {Geim}}]{graphene_membrane_geim}%
  \BibitemOpen
  \bibfield  {author} {\bibinfo {author} {\bibfnamefont {R.~R.}\ \bibnamefont
  {Nair}}, \bibinfo {author} {\bibfnamefont {H.~A.}\ \bibnamefont {Wu}},
  \bibinfo {author} {\bibfnamefont {P.~N.}\ \bibnamefont {Jayaram}}, \bibinfo
  {author} {\bibfnamefont {I.~V.}\ \bibnamefont {Grigorieva}}, \ and\ \bibinfo
  {author} {\bibfnamefont {A.~K.}\ \bibnamefont {Geim}},\ }\href {\doibase
  10.1126/science.1211694} {\bibfield  {journal} {\bibinfo  {journal}
  {Science}\ }\textbf {\bibinfo {volume} {335}},\ \bibinfo {pages} {442}
  (\bibinfo {year} {2012})}\BibitemShut {NoStop}%
\bibitem [{\citenamefont {Logan}\ and\ \citenamefont
  {Elimelech}(2012)}]{Logan_membrane_power_generation}%
  \BibitemOpen
  \bibfield  {author} {\bibinfo {author} {\bibfnamefont {B.~E.}\ \bibnamefont
  {Logan}}\ and\ \bibinfo {author} {\bibfnamefont {M.}~\bibnamefont
  {Elimelech}},\ }\href@noop {} {\bibfield  {journal} {\bibinfo  {journal}
  {Nature}\ }\textbf {\bibinfo {volume} {488}},\ \bibinfo {pages} {313}
  (\bibinfo {year} {2012})}\BibitemShut {NoStop}%
\bibitem [{\citenamefont {Bocquet}\ and\ \citenamefont
  {Charlaix}(2010)}]{Bocquet2010}%
  \BibitemOpen
  \bibfield  {author} {\bibinfo {author} {\bibfnamefont {L.}~\bibnamefont
  {Bocquet}}\ and\ \bibinfo {author} {\bibfnamefont {E.}~\bibnamefont
  {Charlaix}},\ }\href {\doibase 10.1039/b909366b} {\bibfield  {journal}
  {\bibinfo  {journal} {Chem. Soc. Rev.}\ }\textbf {\bibinfo {volume} {39}},\
  \bibinfo {pages} {1073} (\bibinfo {year} {2010})}\BibitemShut {NoStop}%
\bibitem [{\citenamefont {Cohen-Tanugi}\ and\ \citenamefont
  {Grossman}(2012)}]{Desalination_acsnano}%
  \BibitemOpen
  \bibfield  {author} {\bibinfo {author} {\bibfnamefont {D.}~\bibnamefont
  {Cohen-Tanugi}}\ and\ \bibinfo {author} {\bibfnamefont {J.~C.}\ \bibnamefont
  {Grossman}},\ }\href {\doibase 10.1021/nl3012853} {\bibfield  {journal}
  {\bibinfo  {journal} {Nano Letters}\ }\textbf {\bibinfo {volume} {12}},\
  \bibinfo {pages} {3602} (\bibinfo {year} {2012})}\BibitemShut {NoStop}%
\bibitem [{\citenamefont {Konatham}\ \emph {et~al.}(2013)\citenamefont
  {Konatham}, \citenamefont {Yu}, \citenamefont {Ho},\ and\ \citenamefont
  {Striolo}}]{striolo_langmuir}%
  \BibitemOpen
  \bibfield  {author} {\bibinfo {author} {\bibfnamefont {D.}~\bibnamefont
  {Konatham}}, \bibinfo {author} {\bibfnamefont {J.}~\bibnamefont {Yu}},
  \bibinfo {author} {\bibfnamefont {T.~A.}\ \bibnamefont {Ho}}, \ and\ \bibinfo
  {author} {\bibfnamefont {A.}~\bibnamefont {Striolo}},\ }\href@noop {}
  {\bibfield  {journal} {\bibinfo  {journal} {Langmuir}\ }\textbf {\bibinfo
  {volume} {29}},\ \bibinfo {pages} {11884} (\bibinfo {year}
  {2013})}\BibitemShut {NoStop}%
\bibitem [{\citenamefont {Bocquet}\ and\ \citenamefont
  {Barrat}(2007)}]{Bocquet2007}%
  \BibitemOpen
  \bibfield  {author} {\bibinfo {author} {\bibfnamefont {L.}~\bibnamefont
  {Bocquet}}\ and\ \bibinfo {author} {\bibfnamefont {J.~L.}\ \bibnamefont
  {Barrat}},\ }\href@noop {} {\bibfield  {journal} {\bibinfo  {journal} {Soft
  Matter}\ }\textbf {\bibinfo {volume} {3}},\ \bibinfo {pages} {685} (\bibinfo
  {year} {2007})}\BibitemShut {NoStop}%
\bibitem [{\citenamefont {Falk}\ \emph {et~al.}(2010)\citenamefont {Falk},
  \citenamefont {Sedlmeier}, \citenamefont {Joly}, \citenamefont {Netz},\ and\
  \citenamefont {Bocquet}}]{laurent_friction}%
  \BibitemOpen
  \bibfield  {author} {\bibinfo {author} {\bibfnamefont {K.}~\bibnamefont
  {Falk}}, \bibinfo {author} {\bibfnamefont {F.}~\bibnamefont {Sedlmeier}},
  \bibinfo {author} {\bibfnamefont {L.}~\bibnamefont {Joly}}, \bibinfo {author}
  {\bibfnamefont {R.~R.}\ \bibnamefont {Netz}}, \ and\ \bibinfo {author}
  {\bibfnamefont {L.}~\bibnamefont {Bocquet}},\ }\href@noop {} {\bibfield
  {journal} {\bibinfo  {journal} {Nano Lett.}\ }\textbf {\bibinfo {volume}
  {10}},\ \bibinfo {pages} {4067} (\bibinfo {year} {2010})}\BibitemShut
  {NoStop}%
\bibitem [{\citenamefont {Won}\ and\ \citenamefont
  {Aluru}(2007)}]{water_BN_MD_won}%
  \BibitemOpen
  \bibfield  {author} {\bibinfo {author} {\bibfnamefont {C.~Y.}\ \bibnamefont
  {Won}}\ and\ \bibinfo {author} {\bibfnamefont {N.~R.}\ \bibnamefont
  {Aluru}},\ }\href@noop {} {\bibfield  {journal} {\bibinfo  {journal} {{J. Am.
  Chem. Soc.}}\ }\textbf {\bibinfo {volume} {129}},\ \bibinfo {pages} {2748}
  (\bibinfo {year} {2007})}\BibitemShut {NoStop}%
\bibitem [{\citenamefont {Won}\ and\ \citenamefont
  {Aluru}(2008)}]{MD_aluru_jpcc}%
  \BibitemOpen
  \bibfield  {author} {\bibinfo {author} {\bibfnamefont {C.~Y.}\ \bibnamefont
  {Won}}\ and\ \bibinfo {author} {\bibfnamefont {N.~R.}\ \bibnamefont
  {Aluru}},\ }\href@noop {} {\bibfield  {journal} {\bibinfo  {journal} {{J.
  Phys. Chem. C}}\ }\textbf {\bibinfo {volume} {112}},\ \bibinfo {pages} {1182}
  (\bibinfo {year} {2008})}\BibitemShut {NoStop}%
\bibitem [{\citenamefont {Nigu\`{e}s}\ \emph {et~al.}(2014)\citenamefont
  {Nigu\`{e}s}, \citenamefont {Siria}, \citenamefont {Vincent}, \citenamefont
  {P.},\ and\ \citenamefont {Bocquet}}]{BNNT_bouquet_friction}%
  \BibitemOpen
  \bibfield  {author} {\bibinfo {author} {\bibfnamefont {A.}~\bibnamefont
  {Nigu\`{e}s}}, \bibinfo {author} {\bibfnamefont {A.}~\bibnamefont {Siria}},
  \bibinfo {author} {\bibfnamefont {P.}~\bibnamefont {Vincent}}, \bibinfo
  {author} {\bibfnamefont {P.}~\bibnamefont {P.}}, \ and\ \bibinfo {author}
  {\bibfnamefont {L.}~\bibnamefont {Bocquet}},\ }\href@noop {} {\bibfield
  {journal} {\bibinfo  {journal} {{Nat. Mater.}}\ }\textbf {\bibinfo {volume}
  {13}},\ \bibinfo {pages} {688} (\bibinfo {year} {2014})}\BibitemShut
  {NoStop}%
\bibitem [{\citenamefont {Zhang}\ \emph {et~al.}(2013)\citenamefont {Zhang},
  \citenamefont {Ning}, \citenamefont {Y.}, \citenamefont {Q.}, \citenamefont
  {Chen}, \citenamefont {H.}, \citenamefont {Zhang}, \citenamefont {W.},\ and\
  \citenamefont {Wei}}]{super_lubricity_CNT}%
  \BibitemOpen
  \bibfield  {author} {\bibinfo {author} {\bibfnamefont {R.}~\bibnamefont
  {Zhang}}, \bibinfo {author} {\bibfnamefont {Z.}~\bibnamefont {Ning}},
  \bibinfo {author} {\bibfnamefont {Z.}~\bibnamefont {Y.}}, \bibinfo {author}
  {\bibfnamefont {Z.}~\bibnamefont {Q.}}, \bibinfo {author} {\bibfnamefont
  {Q.}~\bibnamefont {Chen}}, \bibinfo {author} {\bibfnamefont {X.}~\bibnamefont
  {H.}}, \bibinfo {author} {\bibfnamefont {Q.}~\bibnamefont {Zhang}}, \bibinfo
  {author} {\bibfnamefont {Q.}~\bibnamefont {W.}}, \ and\ \bibinfo {author}
  {\bibfnamefont {F.}~\bibnamefont {Wei}},\ }\href@noop {} {\bibfield
  {journal} {\bibinfo  {journal} {Nat. Nanotech.}\ }\textbf {\bibinfo {volume}
  {8}},\ \bibinfo {pages} {912} (\bibinfo {year} {2013})}\BibitemShut {NoStop}%
\bibitem [{\citenamefont {Li}\ and\ \citenamefont
  {Zeng}(2012)}]{water_contact_angle_gra_bn}%
  \BibitemOpen
  \bibfield  {author} {\bibinfo {author} {\bibfnamefont {H.}~\bibnamefont
  {Li}}\ and\ \bibinfo {author} {\bibfnamefont {X.~C.}\ \bibnamefont {Zeng}},\
  }\href@noop {} {\bibfield  {journal} {\bibinfo  {journal} {ACS Nano}\
  }\textbf {\bibinfo {volume} {6}},\ \bibinfo {pages} {2401} (\bibinfo {year}
  {2012})}\BibitemShut {NoStop}%
\bibitem [{\citenamefont {Vanossi}\ \emph {et~al.}(2013)\citenamefont
  {Vanossi}, \citenamefont {Manini}, \citenamefont {Urbakh}, \citenamefont
  {Zapperi},\ and\ \citenamefont {Tosatti}}]{tosatti_colloquium}%
  \BibitemOpen
  \bibfield  {author} {\bibinfo {author} {\bibfnamefont {A.}~\bibnamefont
  {Vanossi}}, \bibinfo {author} {\bibfnamefont {N.}~\bibnamefont {Manini}},
  \bibinfo {author} {\bibfnamefont {M.}~\bibnamefont {Urbakh}}, \bibinfo
  {author} {\bibfnamefont {S.}~\bibnamefont {Zapperi}}, \ and\ \bibinfo
  {author} {\bibfnamefont {E.}~\bibnamefont {Tosatti}},\ }\href@noop {}
  {\bibfield  {journal} {\bibinfo  {journal} {Rev. Mod. Phys.}\ }\textbf
  {\bibinfo {volume} {85}},\ \bibinfo {pages} {529} (\bibinfo {year}
  {2013})}\BibitemShut {NoStop}%
\bibitem [{\citenamefont {Shih}\ \emph {et~al.}(2012)\citenamefont {Shih},
  \citenamefont {Wang}, \citenamefont {Lin}, \citenamefont {Park},
  \citenamefont {Jin}, \citenamefont {Strano},\ and\ \citenamefont
  {Blankschtein}}]{wetting_break_down}%
  \BibitemOpen
  \bibfield  {author} {\bibinfo {author} {\bibfnamefont {C.-J.}\ \bibnamefont
  {Shih}}, \bibinfo {author} {\bibfnamefont {Q.~H.}\ \bibnamefont {Wang}},
  \bibinfo {author} {\bibfnamefont {S.}~\bibnamefont {Lin}}, \bibinfo {author}
  {\bibfnamefont {K.-C.}\ \bibnamefont {Park}}, \bibinfo {author}
  {\bibfnamefont {Z.}~\bibnamefont {Jin}}, \bibinfo {author} {\bibfnamefont
  {M.~S.}\ \bibnamefont {Strano}}, \ and\ \bibinfo {author} {\bibfnamefont
  {D.}~\bibnamefont {Blankschtein}},\ }\href@noop {} {\bibfield  {journal}
  {\bibinfo  {journal} {Phys. Rev. Lett.}\ }\textbf {\bibinfo {volume} {109}},\
  \bibinfo {pages} {176101} (\bibinfo {year} {2012})}\BibitemShut {NoStop}%
\bibitem [{\citenamefont {Rafiee}\ \emph {et~al.}(2012)\citenamefont {Rafiee},
  \citenamefont {Mi}, \citenamefont {Gullapalli}, \citenamefont {Thomas},
  \citenamefont {Yavari}, \citenamefont {Shi}, \citenamefont {Ajayan},\ and\
  \citenamefont {Koratkar}}]{wetting_transparency}%
  \BibitemOpen
  \bibfield  {author} {\bibinfo {author} {\bibfnamefont {J.}~\bibnamefont
  {Rafiee}}, \bibinfo {author} {\bibfnamefont {X.}~\bibnamefont {Mi}}, \bibinfo
  {author} {\bibfnamefont {H.}~\bibnamefont {Gullapalli}}, \bibinfo {author}
  {\bibfnamefont {A.~V.}\ \bibnamefont {Thomas}}, \bibinfo {author}
  {\bibfnamefont {F.}~\bibnamefont {Yavari}}, \bibinfo {author} {\bibfnamefont
  {Y.}~\bibnamefont {Shi}}, \bibinfo {author} {\bibfnamefont {P.~M.}\
  \bibnamefont {Ajayan}}, \ and\ \bibinfo {author} {\bibfnamefont {N.~A.}\
  \bibnamefont {Koratkar}},\ }\href@noop {} {\bibfield  {journal} {\bibinfo
  {journal} {Nat. Mater.}\ }\textbf {\bibinfo {volume} {11}},\ \bibinfo {pages}
  {217} (\bibinfo {year} {2012})}\BibitemShut {NoStop}%
\bibitem [{\citenamefont {Cicero}\ \emph {et~al.}(2008)\citenamefont {Cicero},
  \citenamefont {Grossman}, \citenamefont {Schwegler}, \citenamefont {Gygi},\
  and\ \citenamefont {Galli}}]{Galli_water_graphene}%
  \BibitemOpen
  \bibfield  {author} {\bibinfo {author} {\bibfnamefont {G.}~\bibnamefont
  {Cicero}}, \bibinfo {author} {\bibfnamefont {J.~C.}\ \bibnamefont
  {Grossman}}, \bibinfo {author} {\bibfnamefont {E.}~\bibnamefont {Schwegler}},
  \bibinfo {author} {\bibfnamefont {F.}~\bibnamefont {Gygi}}, \ and\ \bibinfo
  {author} {\bibfnamefont {G.}~\bibnamefont {Galli}},\ }\href@noop {}
  {\bibfield  {journal} {\bibinfo  {journal} {{J. Am. Chem. Soc.}}\ }\textbf
  {\bibinfo {volume} {130}},\ \bibinfo {pages} {1871} (\bibinfo {year}
  {2008})}\BibitemShut {NoStop}%
\bibitem [{\citenamefont {Tocci}\ and\ \citenamefont
  {Michaelides}(2014)}]{tocci2014ZnO}%
  \BibitemOpen
  \bibfield  {author} {\bibinfo {author} {\bibfnamefont {G.}~\bibnamefont
  {Tocci}}\ and\ \bibinfo {author} {\bibfnamefont {A.}~\bibnamefont
  {Michaelides}},\ }\href {\doibase 10.1021/jz402646c} {\bibfield  {journal}
  {\bibinfo  {journal} {J. Phys. Chem. Lett.}\ }\textbf {\bibinfo {volume}
  {5}},\ \bibinfo {pages} {474} (\bibinfo {year} {2014})}\BibitemShut {NoStop}%
\bibitem [{\citenamefont {K\"{u}hne}\ \emph {et~al.}(1998)\citenamefont
  {K\"{u}hne}, \citenamefont {Krack},\ and\ \citenamefont
  {Parrinello}}]{kuhne_viscosity}%
  \BibitemOpen
  \bibfield  {author} {\bibinfo {author} {\bibnamefont {K\"{u}hne}}, \bibinfo
  {author} {\bibfnamefont {M.}~\bibnamefont {Krack}}, \ and\ \bibinfo {author}
  {\bibfnamefont {M.}~\bibnamefont {Parrinello}},\ }\href@noop {} {\bibfield
  {journal} {\bibinfo  {journal} {{J. Chem. Theor. Comput.}}\ }\textbf
  {\bibinfo {volume} {23}},\ \bibinfo {pages} {5161} (\bibinfo {year}
  {1998})}\BibitemShut {NoStop}%
\bibitem [{\citenamefont {Alf\`{e}}\ and\ \citenamefont
  {Gillan}(1998)}]{alfe_viscosity}%
  \BibitemOpen
  \bibfield  {author} {\bibinfo {author} {\bibfnamefont {D.}~\bibnamefont
  {Alf\`{e}}}\ and\ \bibinfo {author} {\bibfnamefont {M.~J.}\ \bibnamefont
  {Gillan}},\ }\href@noop {} {\bibfield  {journal} {\bibinfo  {journal} {{Phys.
  Rev. Lett.}}\ }\textbf {\bibinfo {volume} {23}},\ \bibinfo {pages} {5161}
  (\bibinfo {year} {1998})}\BibitemShut {NoStop}%
\bibitem [{\citenamefont {Feibelman}(2013)}]{feibelman_viscosity}%
  \BibitemOpen
  \bibfield  {author} {\bibinfo {author} {\bibfnamefont {P.~J.}\ \bibnamefont
  {Feibelman}},\ }\href {\doibase 10.1021/jp312152h} {\bibfield  {journal}
  {\bibinfo  {journal} {J. Phys. Chem. C}\ }\textbf {\bibinfo {volume} {117}},\
  \bibinfo {pages} {6088} (\bibinfo {year} {2013})}\BibitemShut {NoStop}%
\bibitem [{not()}]{note_graphene}%
  \BibitemOpen
  \href@noop {} {}\bibinfo {note} {The effect of nanotube curvature on the
  dynamics becomes negligible for a tube radius $R > 10$
  nm.~\cite{laurent_friction,Falk2012}}\BibitemShut {NoStop}%
\bibitem [{\citenamefont {{J. VandeVondele and M. Krack and F. Mohamed and M.
  Parrinello and T. Chassaing and J. Hutter}}(2005)}]{cp2kvond}%
  \BibitemOpen
  \bibfield  {author} {\bibinfo {author} {\bibnamefont {{J. VandeVondele and M.
  Krack and F. Mohamed and M. Parrinello and T. Chassaing and J. Hutter}}},\
  }\href@noop {} {\bibfield  {journal} {\bibinfo  {journal} {{Comp. Phys.
  Comm.}}\ }\textbf {\bibinfo {volume} {167}},\ \bibinfo {pages} {103}
  (\bibinfo {year} {2005})}\BibitemShut {NoStop}%
\bibitem [{\citenamefont {Klime\v{s}}\ \emph {et~al.}(2010)\citenamefont
  {Klime\v{s}}, \citenamefont {Bowler},\ and\ \citenamefont
  {Michaelides}}]{jiri_molecules}%
  \BibitemOpen
  \bibfield  {author} {\bibinfo {author} {\bibfnamefont {J.}~\bibnamefont
  {Klime\v{s}}}, \bibinfo {author} {\bibfnamefont {D.~R.}\ \bibnamefont
  {Bowler}}, \ and\ \bibinfo {author} {\bibfnamefont {A.}~\bibnamefont
  {Michaelides}},\ }\href@noop {} {\bibfield  {journal} {\bibinfo  {journal}
  {{J. Phys.: Condens. Matter}}\ }\textbf {\bibinfo {volume} {22}},\ \bibinfo
  {pages} {022201} (\bibinfo {year} {2010})}\BibitemShut {NoStop}%
\bibitem [{\citenamefont {Klime\v{s}}\ \emph {et~al.}(2011)\citenamefont
  {Klime\v{s}}, \citenamefont {Bowler},\ and\ \citenamefont
  {Michaelides}}]{jiri_solids}%
  \BibitemOpen
  \bibfield  {author} {\bibinfo {author} {\bibfnamefont {J.}~\bibnamefont
  {Klime\v{s}}}, \bibinfo {author} {\bibfnamefont {D.~R.}\ \bibnamefont
  {Bowler}}, \ and\ \bibinfo {author} {\bibfnamefont {A.}~\bibnamefont
  {Michaelides}},\ }\href@noop {} {\bibfield  {journal} {\bibinfo  {journal}
  {{Phys. Rev. B}}\ }\textbf {\bibinfo {volume} {83}},\ \bibinfo {pages}
  {195131} (\bibinfo {year} {2011})}\BibitemShut {NoStop}%
\bibitem [{\citenamefont {Graziano}\ \emph {et~al.}(2012)\citenamefont
  {Graziano}, \citenamefont {Klime\v{s}}, \citenamefont {Fernandez-Alonso},\
  and\ \citenamefont {Michaelides}}]{graziano_vdw_gra_bn}%
  \BibitemOpen
  \bibfield  {author} {\bibinfo {author} {\bibfnamefont {G.}~\bibnamefont
  {Graziano}}, \bibinfo {author} {\bibfnamefont {J.}~\bibnamefont
  {Klime\v{s}}}, \bibinfo {author} {\bibfnamefont {F.}~\bibnamefont
  {Fernandez-Alonso}}, \ and\ \bibinfo {author} {\bibfnamefont
  {A.}~\bibnamefont {Michaelides}},\ }\href@noop {} {\bibfield  {journal}
  {\bibinfo  {journal} {{J. Phys.: Condens. Matter}}\ }\textbf {\bibinfo
  {volume} {24}},\ \bibinfo {pages} {424216} (\bibinfo {year}
  {2012})}\BibitemShut {NoStop}%
\bibitem [{\citenamefont {Zhang}\ \emph {et~al.}(2011)\citenamefont {Zhang},
  \citenamefont {Wu}, \citenamefont {Galli},\ and\ \citenamefont
  {Gygi}}]{Galli_water_jctc}%
  \BibitemOpen
  \bibfield  {author} {\bibinfo {author} {\bibfnamefont {C.}~\bibnamefont
  {Zhang}}, \bibinfo {author} {\bibfnamefont {J.}~\bibnamefont {Wu}}, \bibinfo
  {author} {\bibfnamefont {G.}~\bibnamefont {Galli}}, \ and\ \bibinfo {author}
  {\bibfnamefont {F.}~\bibnamefont {Gygi}},\ }\href@noop {} {\bibfield
  {journal} {\bibinfo  {journal} {J. Chem. Theory Comput.}\ }\textbf {\bibinfo
  {volume} {7}},\ \bibinfo {pages} {3054} (\bibinfo {year} {2011})}\BibitemShut
  {NoStop}%
\bibitem [{\citenamefont {Carrasco}\ \emph {et~al.}(2011)\citenamefont
  {Carrasco}, \citenamefont {Santra}, \citenamefont {Klime\v{s}},\ and\
  \citenamefont {Michaelides}}]{javi_vdw_prl}%
  \BibitemOpen
  \bibfield  {author} {\bibinfo {author} {\bibfnamefont {J.}~\bibnamefont
  {Carrasco}}, \bibinfo {author} {\bibfnamefont {B.}~\bibnamefont {Santra}},
  \bibinfo {author} {\bibfnamefont {J.}~\bibnamefont {Klime\v{s}}}, \ and\
  \bibinfo {author} {\bibfnamefont {A.}~\bibnamefont {Michaelides}},\
  }\href@noop {} {\bibfield  {journal} {\bibinfo  {journal} {{Phys. Rev.
  Lett.}}\ }\textbf {\bibinfo {volume} {106}},\ \bibinfo {pages} {026101}
  (\bibinfo {year} {2011})}\BibitemShut {NoStop}%
\bibitem [{\citenamefont {Luzar}\ and\ \citenamefont
  {Chandler}(1996)}]{luzar_chand}%
  \BibitemOpen
  \bibfield  {author} {\bibinfo {author} {\bibfnamefont {A.}~\bibnamefont
  {Luzar}}\ and\ \bibinfo {author} {\bibfnamefont {D.}~\bibnamefont
  {Chandler}},\ }\href@noop {} {\bibfield  {journal} {\bibinfo  {journal}
  {{Nature}}\ }\textbf {\bibinfo {volume} {379}},\ \bibinfo {pages} {55}
  (\bibinfo {year} {1996})}\BibitemShut {NoStop}%
\bibitem [{\citenamefont {Lee}\ \emph {et~al.}(1984)\citenamefont {Lee},
  \citenamefont {McCammon},\ and\ \citenamefont
  {Rossky}}]{liq_wat_surfaces_rossky}%
  \BibitemOpen
  \bibfield  {author} {\bibinfo {author} {\bibfnamefont {C.-Y.}\ \bibnamefont
  {Lee}}, \bibinfo {author} {\bibfnamefont {J.~A.}\ \bibnamefont {McCammon}}, \
  and\ \bibinfo {author} {\bibfnamefont {P.~J.}\ \bibnamefont {Rossky}},\
  }\href@noop {} {\bibfield  {journal} {\bibinfo  {journal} {{J. Chem. Phys.}}\
  }\textbf {\bibinfo {volume} {80}},\ \bibinfo {pages} {4448} (\bibinfo {year}
  {1984})}\BibitemShut {NoStop}%
\bibitem [{\citenamefont {{Ho, Tuan A. and Striolo,
  Alberto}}(2013)}]{striolo_water_graphene_polarizability}%
  \BibitemOpen
  \bibfield  {author} {\bibinfo {author} {\bibnamefont {{Ho, Tuan A. and
  Striolo, Alberto}}},\ }\href {\doibase {http://dx.doi.org/10.1063/1.4789583}}
  {\bibfield  {journal} {\bibinfo  {journal} {{J. Chem. Phys.}}\ }\textbf
  {\bibinfo {volume} {{138}}},\ \bibinfo {pages} {{054117}} (\bibinfo {year}
  {{2013}})}\BibitemShut {NoStop}%
\bibitem [{\citenamefont {Kuo}\ and\ \citenamefont
  {Mundy}(2004)}]{wat_kuo_sci}%
  \BibitemOpen
  \bibfield  {author} {\bibinfo {author} {\bibfnamefont {I.~F.~W.}\
  \bibnamefont {Kuo}}\ and\ \bibinfo {author} {\bibfnamefont {C.~J.}\
  \bibnamefont {Mundy}},\ }\href@noop {} {\bibfield  {journal} {\bibinfo
  {journal} {{Science}}\ }\textbf {\bibinfo {volume} {303}},\ \bibinfo {pages}
  {658} (\bibinfo {year} {2004})}\BibitemShut {NoStop}%
\bibitem [{\citenamefont {Bocquet}\ and\ \citenamefont
  {Barrat}(1994)}]{Bocquet1994}%
  \BibitemOpen
  \bibfield  {author} {\bibinfo {author} {\bibfnamefont {L.}~\bibnamefont
  {Bocquet}}\ and\ \bibinfo {author} {\bibfnamefont {J.~L.}\ \bibnamefont
  {Barrat}},\ }\href@noop {} {\bibfield  {journal} {\bibinfo  {journal} {Phys.
  Rev. E}\ }\textbf {\bibinfo {volume} {49}},\ \bibinfo {pages} {3079}
  (\bibinfo {year} {1994})}\BibitemShut {NoStop}%
\bibitem [{\citenamefont {Bocquet}\ and\ \citenamefont
  {Barrat}(2013)}]{Bocquet2013}%
  \BibitemOpen
  \bibfield  {author} {\bibinfo {author} {\bibfnamefont {L.}~\bibnamefont
  {Bocquet}}\ and\ \bibinfo {author} {\bibfnamefont {J.~L.}\ \bibnamefont
  {Barrat}},\ }\href@noop {} {\bibfield  {journal} {\bibinfo  {journal} {J.
  Chem. Phys.}\ }\textbf {\bibinfo {volume} {139}},\ \bibinfo {pages} {044704}
  (\bibinfo {year} {2013})}\BibitemShut {NoStop}%
\bibitem [{\citenamefont {Huang}\ \emph {et~al.}(2008)\citenamefont {Huang},
  \citenamefont {Sendner}, \citenamefont {Horinek}, \citenamefont {Netz},\ and\
  \citenamefont {Bocquet}}]{Huang2008b}%
  \BibitemOpen
  \bibfield  {author} {\bibinfo {author} {\bibfnamefont {D.~M.}\ \bibnamefont
  {Huang}}, \bibinfo {author} {\bibfnamefont {C.}~\bibnamefont {Sendner}},
  \bibinfo {author} {\bibfnamefont {D.}~\bibnamefont {Horinek}}, \bibinfo
  {author} {\bibfnamefont {R.~R.}\ \bibnamefont {Netz}}, \ and\ \bibinfo
  {author} {\bibfnamefont {L.}~\bibnamefont {Bocquet}},\ }\href@noop {}
  {\bibfield  {journal} {\bibinfo  {journal} {Phys. Rev. Lett.}\ }\textbf
  {\bibinfo {volume} {101}},\ \bibinfo {pages} {226101} (\bibinfo {year}
  {2008})}\BibitemShut {NoStop}%
\bibitem [{\citenamefont {Ortiz-Young}\ \emph {et~al.}(2013)\citenamefont
  {Ortiz-Young}, \citenamefont {Chiu}, \citenamefont {Kim}, \citenamefont
  {Vo\"{i}tchovsky},\ and\ \citenamefont {Riedo}}]{ncomm_young_friction}%
  \BibitemOpen
  \bibfield  {author} {\bibinfo {author} {\bibfnamefont {D.}~\bibnamefont
  {Ortiz-Young}}, \bibinfo {author} {\bibfnamefont {H.-C.}\ \bibnamefont
  {Chiu}}, \bibinfo {author} {\bibfnamefont {S.}~\bibnamefont {Kim}}, \bibinfo
  {author} {\bibfnamefont {K.}~\bibnamefont {Vo\"{i}tchovsky}}, \ and\ \bibinfo
  {author} {\bibfnamefont {E.}~\bibnamefont {Riedo}},\ }\href@noop {}
  {\bibfield  {journal} {\bibinfo  {journal} {Nat. Commun.}\ }\textbf {\bibinfo
  {volume} {4}},\ \bibinfo {pages} {24282} (\bibinfo {year}
  {2013})}\BibitemShut {NoStop}%
\bibitem [{\citenamefont {Ho}\ \emph {et~al.}(2011)\citenamefont {Ho},
  \citenamefont {Papavassiliou}, \citenamefont {Lee},\ and\ \citenamefont
  {Striolo}}]{striolo_PNAS}%
  \BibitemOpen
  \bibfield  {author} {\bibinfo {author} {\bibfnamefont {T.~A.}\ \bibnamefont
  {Ho}}, \bibinfo {author} {\bibfnamefont {D.~V.}\ \bibnamefont
  {Papavassiliou}}, \bibinfo {author} {\bibfnamefont {L.~L.}\ \bibnamefont
  {Lee}}, \ and\ \bibinfo {author} {\bibfnamefont {A.}~\bibnamefont
  {Striolo}},\ }\href {\doibase 10.1073/pnas.1105189108} {\bibfield  {journal}
  {\bibinfo  {journal} {Proc. Natl. Acad. Sci. USA}\ }\textbf {\bibinfo
  {volume} {108}},\ \bibinfo {pages} {16170} (\bibinfo {year}
  {2011})}\BibitemShut {NoStop}%
\bibitem [{\citenamefont {Krim}(2012)}]{krim_friction_rev}%
  \BibitemOpen
  \bibfield  {author} {\bibinfo {author} {\bibfnamefont {J.}~\bibnamefont
  {Krim}},\ }\href {\doibase 10.1080/00018732.2012.706401} {\bibfield
  {journal} {\bibinfo  {journal} {Adv. Phys.}\ }\textbf {\bibinfo {volume}
  {61}},\ \bibinfo {pages} {155} (\bibinfo {year} {2012})}\BibitemShut
  {NoStop}%
\bibitem [{\citenamefont {Maali}\ \emph {et~al.}(2008)\citenamefont {Maali},
  \citenamefont {Cohen-Bouhacina},\ and\ \citenamefont
  {Kellay}}]{mali_friction_water_graphite}%
  \BibitemOpen
  \bibfield  {author} {\bibinfo {author} {\bibfnamefont {A.}~\bibnamefont
  {Maali}}, \bibinfo {author} {\bibfnamefont {T.}~\bibnamefont
  {Cohen-Bouhacina}}, \ and\ \bibinfo {author} {\bibfnamefont {H.}~\bibnamefont
  {Kellay}},\ }\href@noop {} {\bibfield  {journal} {\bibinfo  {journal} {Appl.
  Phys. Lett.}\ }\textbf {\bibinfo {volume} {92}},\ \bibinfo {pages} {053101}
  (\bibinfo {year} {2008})}\BibitemShut {NoStop}%
\bibitem [{\citenamefont {Gobre}\ and\ \citenamefont
  {Tkatchenko}(2013)}]{tkatc_natur_comm_2013}%
  \BibitemOpen
  \bibfield  {author} {\bibinfo {author} {\bibfnamefont {V.~V.}\ \bibnamefont
  {Gobre}}\ and\ \bibinfo {author} {\bibfnamefont {A.}~\bibnamefont
  {Tkatchenko}},\ }\href@noop {} {\bibfield  {journal} {\bibinfo  {journal}
  {{Nat. Commun.}}\ }\textbf {\bibinfo {volume} {4}},\ \bibinfo {pages} {2341}
  (\bibinfo {year} {2013})}\BibitemShut {NoStop}%
\bibitem [{\citenamefont {{Kannam, Sridhar Kumar and Todd, B. D. and Hansen, J.
  S. and Daivis, Peter J.}}(2013)}]{Kannam_JCP}%
  \BibitemOpen
  \bibfield  {author} {\bibinfo {author} {\bibnamefont {{Kannam, Sridhar Kumar
  and Todd, B. D. and Hansen, J. S. and Daivis, Peter J.}}},\ }\href {\doibase
  http://dx.doi.org/10.1063/1.4793396} {\bibfield  {journal} {\bibinfo
  {journal} {J. Chem. Phys.}\ }\textbf {\bibinfo {volume} {138}},\ \bibinfo
  {pages} {094701} (\bibinfo {year} {2013})}\BibitemShut {NoStop}%
\bibitem [{\citenamefont {Werder}\ \emph {et~al.}(2003)\citenamefont {Werder},
  \citenamefont {Walther}, \citenamefont {Jaffe}, \citenamefont {Halicioglu},\
  and\ \citenamefont {Koumoutsakos}}]{Werder2003}%
  \BibitemOpen
  \bibfield  {author} {\bibinfo {author} {\bibfnamefont {T.}~\bibnamefont
  {Werder}}, \bibinfo {author} {\bibfnamefont {J.~H.}\ \bibnamefont {Walther}},
  \bibinfo {author} {\bibfnamefont {R.~L.}\ \bibnamefont {Jaffe}}, \bibinfo
  {author} {\bibfnamefont {T.}~\bibnamefont {Halicioglu}}, \ and\ \bibinfo
  {author} {\bibfnamefont {P.}~\bibnamefont {Koumoutsakos}},\ }\href@noop {}
  {\bibfield  {journal} {\bibinfo  {journal} {{J. Phys. Chem. B}}\ }\textbf
  {\bibinfo {volume} {107}},\ \bibinfo {pages} {1345} (\bibinfo {year}
  {2003})}\BibitemShut {NoStop}%
\bibitem [{\citenamefont {{J. Ma and A. Michaelides and D. Alf\`{e} and L
  Schimka and G. Kresse and E. Wang }}(2011)}]{ma_monomer_on_gra}%
  \BibitemOpen
  \bibfield  {author} {\bibinfo {author} {\bibnamefont {{J. Ma and A.
  Michaelides and D. Alf\`{e} and L Schimka and G. Kresse and E. Wang }}},\
  }\href@noop {} {\bibfield  {journal} {\bibinfo  {journal} {{Phys. Rev. B}}\
  }\textbf {\bibinfo {volume} {84}},\ \bibinfo {pages} {003402} (\bibinfo
  {year} {2011})}\BibitemShut {NoStop}%
\bibitem [{\citenamefont {Geim}\ and\ \citenamefont
  {Grigorieva}(2013)}]{vdw_hetero_structures}%
  \BibitemOpen
  \bibfield  {author} {\bibinfo {author} {\bibfnamefont {A.~K.}\ \bibnamefont
  {Geim}}\ and\ \bibinfo {author} {\bibfnamefont {I.~V.}\ \bibnamefont
  {Grigorieva}},\ }\href@noop {} {\bibfield  {journal} {\bibinfo  {journal}
  {Nature}\ }\textbf {\bibinfo {volume} {499}},\ \bibinfo {pages} {419}
  (\bibinfo {year} {2013})}\BibitemShut {NoStop}%
\bibitem [{\citenamefont {Ma}\ \emph {et~al.}(2011)\citenamefont {Ma},
  \citenamefont {Shen}, \citenamefont {Sheridan}, \citenamefont {Liu},
  \citenamefont {Chen},\ and\ \citenamefont {Zheng}}]{ming_ma_PRE_2011}%
  \BibitemOpen
  \bibfield  {author} {\bibinfo {author} {\bibfnamefont {M.~D.}\ \bibnamefont
  {Ma}}, \bibinfo {author} {\bibfnamefont {L.}~\bibnamefont {Shen}}, \bibinfo
  {author} {\bibfnamefont {J.}~\bibnamefont {Sheridan}}, \bibinfo {author}
  {\bibfnamefont {J.~Z.}\ \bibnamefont {Liu}}, \bibinfo {author} {\bibfnamefont
  {C.}~\bibnamefont {Chen}}, \ and\ \bibinfo {author} {\bibfnamefont
  {Q.}~\bibnamefont {Zheng}},\ }\href {\doibase 10.1103/PhysRevE.83.036316}
  {\bibfield  {journal} {\bibinfo  {journal} {Phys. Rev. E}\ }\textbf {\bibinfo
  {volume} {83}},\ \bibinfo {pages} {036316} (\bibinfo {year}
  {2011})}\BibitemShut {NoStop}%
\bibitem [{\citenamefont {Falk}\ \emph {et~al.}(2012)\citenamefont {Falk},
  \citenamefont {Sedlmeier}, \citenamefont {Joly}, \citenamefont {Netz},\ and\
  \citenamefont {Bocquet}}]{Falk2012}%
  \BibitemOpen
  \bibfield  {author} {\bibinfo {author} {\bibfnamefont {K.}~\bibnamefont
  {Falk}}, \bibinfo {author} {\bibfnamefont {F.}~\bibnamefont {Sedlmeier}},
  \bibinfo {author} {\bibfnamefont {L.}~\bibnamefont {Joly}}, \bibinfo {author}
  {\bibfnamefont {R.~R.}\ \bibnamefont {Netz}}, \ and\ \bibinfo {author}
  {\bibfnamefont {L.}~\bibnamefont {Bocquet}},\ }\href@noop {} {\bibfield
  {journal} {\bibinfo  {journal} {Langmuir}\ }\textbf {\bibinfo {volume}
  {28}},\ \bibinfo {pages} {14261} (\bibinfo {year} {2012})}\BibitemShut
  {NoStop}%
\end{thebibliography}%
\end{document}